\documentclass{jpconf}

\usepackage{amssymb}

\newtheorem{theorem}{Theorem}

\newtheorem{remark}[theorem]{Remark}

\usepackage{iopams}

\begin{document}
\title{\large Presented at \\Physics Beyond Relativity 2019 conference \\ Prague, Czech Republic, October 20, 2019
\\ (invited presentation)  \\ \vspace{2cm} \huge Logical Analysis of Relativity Theory}

\author{Akira Kanda$^1$, Mihai Prunescu$^2$ and Renata Wong$^3$}

\address{$^1$ Omega Mathematical Institute/University of Toronto, Toronto, Ontario, Canada}
\address{$^2$ University of Bucharest, Bucharest, Romania} 
\address{$^3$ Department of Computer Science and Technology, Nanjing University, Nanjing, China}

\ead{kanda@cs.toronto.edu, mihai.prunescu@gmail.com, renata.wong@protonmail.com}

\section{Prelude}

\subsection{Newton v.s. Galileo: Reclining Tower Experiment}

\textrm{Through analysing Kepler's massive data on the observed motion of
planets, Newton reached the conclusion that our planets are orbiting around
the Sun and the cause of such circular motion is the gravitational pull
(centripetal force) exerted by the Sun on the planets. Upon this discovery,
Newton developed the first theory of dynamics in human history. \textsl{This
should be understood as the first major example of how empiricism
contributed to building a most important theory of physics}. }

\textrm{Interestingly, this success also gives a stern warning to the
popularized thesis that this process is one-way only. Indeed, as we will show in
what follows, that\textsl{\ Newton's dynamics refutes Galileo's famous
reclining tower experiment. }}

\textrm{Since the famous reclining tower experiment of Galileo, students of
physics have been told that contrary to the ancient claim of Aristotle, the
speed of a falling light object and that of a heavy object were the same when dropped from the same height. }

\textrm{In what follows, we will see that, according to Newtonian mechanics, 
\textsl{Aristotle was correct,} namely that the heavier the mass the faster it falls. }

\textrm{Assume }$m$\textrm{\ and }$m\prime $\textrm{\ are two masses with }$%
m>m\prime $\textrm{. Let }$r$\textrm{\ be the distance from the centre of
the Earth to both }$m$\textrm{\ and }$m\prime $\textrm{. Let }$M$\textrm{\ be the
mass of the Earth. The gravitational force between the Earth and }$m$\textrm{\
is }$F=GMm/r^{2}.$\textrm{\ So, }%
\[
a_{m}=GM/r^{2},\quad a_{M}=Gm/r^{2} 
\]%
\textrm{where }$a_{m}$\textrm{\ is the absolute acceleration of }$m$\textrm{%
\ due to the gravitational pull by }$M$\textrm{. The absolute acceleration
for }$M$\textrm{\ in the absolute space is }$-a_{M}$\textrm{. Therefore the
relative acceleration between }$M$\textrm{\ and }$m$\textrm{\ is }$%
a_{M}+a_{m}=GM/r^{2}+Gm/r^{2}.$
\textrm{The time required for }$m$\textrm{\ and }$m\prime $\textrm{\ to
reach }$M$\textrm{\ is not the same unless }$m=m\prime $\textrm{. If }$%
m>m\prime $\textrm{, we have }%
\[
t_{m\prime }>t_{m}. 
\]%
\textrm{This means ``\textsl{The heavier the mass the faster it falls.}"}

\textrm{This is a clear example of \textsl{how inaccurate experiment could
readily lead us to the wrong conclusion.} Physics community believed in this
wrong prediction by Galileo for four centuries until we brought up this
problem just a few years ago. }

\begin{remark}
\textrm{Indeed, in a more advanced stage of quantum mechanics, Kuhn pointed
out that there is no such thing as the so-called experimental verification
or even refutation. This is because the prediction in quantum mechanics is intrinsically
probabilistic and the probability theory states that the relative frequency
converges only at the limit. Indeed, one cannot experimentally prove that in tossing a coin the probability of obtaining heads is 0.5.}
\end{remark}

\subsection{Newton's absolutism v.s. Galileo's relativism: God
v.s. human}

\textrm{Galileo assumed that each observer in this universe has his/her own
reference frame in which physical phenomena take place. He further assumed that the
observer is stationary in his/her own reference frame and that some objects are
stationary in his/her frame while others are in motion in his/her frame. This means
that the entire frame of another observer moving in his frame is moving
inside his frame.}

\textrm{All of this makes the observed phenomena appear different in different reference
frames. To resolve this problem, Galileo assumed that every
reference frame has its own independent status and that reference frames are connected to
each other through the so-called  \textsl{Principle of Relativity} which
asserts that all valid laws of physics must be shared by all reference
frames.}

\textrm{Newton objected to this view of Galileo and presented a strikingly
different view of relativism (relative motion). Within the absolute
theory of dynamics as discussed above, Newton defined ``relative motion" as 
\textsl{the difference between two absolute motions in the absolute frame of
the universe}. So, unlike Galileo's view, \textsl{relative motions are to be
observed only by an observer standing on the shoulder of God from outside of
the physical universe.} For Galileo, the relative motions are to be observed
by the moving observer. }

\textrm{Without this absolute perspective, Galileo's relativity theory
suffered from a fundamental obscurity. It has always been a controversial issue
in relativity theory as to whether if A observes B moving with speed }$v$\textrm{, B should observe that A is moving with speed} $-v$\textrm{. Some relativists agree and some disagree. \textsl{After all, assuming this is violating
the most important assumption of ``symmetry" which is essential to relativism.}
Why a difference in the sign? To begin with, this issue raises a fundamental
question on the validity of Galilean relativity theory and its extension,
Einsteinian relativity theory. This certainly gives rise to a most
fundamental question of what do we mean by \textsl{``observation"}. As
an ``empiricism", relativism has a duty to make this point clear. }

\textrm{In the fast developing phase of the theoretical physics this
fundamental difference between these views of the two giants have
been overlooked. A major purpose of this monograph is to make the relation
between them articulate so as to clear the existing confusion.}

\textrm{Galileo had not articulated laws of physics as axioms}\textsl{.}%
\textrm{\ He experimentally proved that regardless of the magnitude of
``masses", two masses take the same time to reach the ground when released
from the same height. Naturally, he had no theory to precisely prove this
experimental fact. \textsl{So,} \textsl{it was impossible for his
contemporaries to evaluate the validity (or the effectiveness) of the
principle of relativity.} A century later, Newton's dynamics presented
legitimate laws of physics and it was discovered that when we consider two
Galilean reference frames that are in acceleration with respect to each
other the principle of relativity fails. The standard argument goes as
follows: Assume a train is in acceleration on a track. On the embankment,
there is a tree. An observer on the embankment will observe that the train
is under force. Another observer on the embankment will observe that the
tree is under acceleration. So, Post-Newtonian era Galilean relativists
resolved this problem \textsl{by rejecting reference frames under
acceleration}. In this way they abandoned most of Newton's dynamics.
From this point on, researchers stopped paying much attention to this
problem until it resurfaced in the late 19th century when the 
Michelson-Morley experiment threw physics into a serious crisis from which
Einstein's special theory of relativity emerged.}

\textrm{What is not clear is whether this argument involves the  applicability of the Newtonian 
laws of physics to Galilean relativity theory. Newton assumed
these laws only for the absolute frame. This problem can be more clearly
explicated as follows: Assume we allow moving reference frames inside
Newton's absolute frame. Assume }$m$\textrm{\ and }$M$\textrm{\ are
attracting each other with the gravitational force. If we consider a
reference frame for }$M$\textrm{\ that moves inside the absolute frame of
Newton, we end up with violating the third law of Newton's as }$M$\textrm{\ is
not moving in this frame. So, there is no hope for expecting Newton's law of
physics to be shared by all reference frames.}

\textrm{The Newtonian interpretation above is based upon the observation
from the outside of the universe. Galileo was observing it from the frame of
himself and the reclining tower, which is the frame of }$M$\textrm{. This
will give us an entirely different conclusion. According to Galileo, }$M$%
\textrm{\ does not move and so the acceleration }$a_{m}$\textrm{\ between }$%
m $\textrm{\ and }$M$\textrm{\ is }$GM/r^{2}.$\textrm{\ Similarly, the
acceleration }$a_{m^{\prime }}$\textrm{\ between }$m^{\prime }$\textrm{\ and 
}$M$\textrm{\ is }$GM/r^{2}.$\textrm{\ So both }$m$\textrm{\ and }$m\prime $%
\textrm{\ will reach the ground within the same time interval. This result
is false as the reasoning behind it ignored the action-reaction law in between }$m$%
\textrm{\ and }$M$\textrm{. In classical dynamics we accept the assumption
of point mass. }

\textrm{This deep philosophical problem will resurface in the modern setting
when Einsteinian relativists present the light bend observation as the
empirical verification of the claim of the general theory of relativity. }

\subsection{Classical Electromagnetism}

\textrm{The theory of electromagnetism was started by Coulomb as an action at a
distance theory of electromagnetic forces. The first systematic theorization
was done by Gauss and Weber. As Faraday's work, such as Faraday induction,
made the theory more complex than Newton's gravitational theory, Heaviside
and Hertz moved towards a theory that was based upon electromagnetic
force field diverting from the action at a distance theory. Having
been influenced by Heaviside and Hertz, Maxwell reluctantly formalized the electromagnetic
theory as a field theory. However, together with Lorentz, Maxwell was trying
to deploy an action-reaction-based interpretation of the field-based theory to
compensate for some fundamental deficiencies of the electromagnetic field
theory. This is known as ``electromagnetic aether theory". Maxwell's electromagnetic field
equations are }%
\[
\mathbf{\nabla \times E}=-\frac{1}{c}\frac{\partial \mathbf{H}}{\partial t}%
,\quad \mathbf{\nabla \cdot E}=4\pi \rho ,\quad \mathbf{\nabla \cdot H}%
=0,\quad \mathbf{J}=\rho \mathbf{v,\quad \nabla \times H}=\frac{4\pi }{c}%
\mathbf{J}+\frac{1}{c}\frac{\partial }{\partial t}\mathbf{E} 
\]%
\textrm{Here }$\mathbf{J}$\textrm{\ is the current vector that transcends
electromagnetic force field and }$\rho $ \textrm{is the charge density of the current. }

\textrm{The first major impact of the introduction of the force field
concept into the formulation of the electromagnetic theory was the apparent increase of the
mass of a charge in the field in terms of acceleration. This problem
already appeared in the fluid dynamics of Stokes.}

\textrm{(1)} \textrm{J. J. Thomson discovered that ``energy of
electromagnetic field" }$E_{em}$ \textrm{gives raise to an additional mass 
}$m=4E_{em}/3c^{2}$\textrm{\ to charged bodies and called it \emph{%
electromagnetic mass (renormalized mass).} }

\textrm{(2) Using the ``momentum of electromagnetic fields",
 Poincar\'{e} showed that these fields contribute an additional mass }$E_{em}/c^{2}$%
\textrm{\ to a charged body. }

\textrm{(3)} \textrm{J.J}. \textrm{Thomson further observed that
electromagnetic mass (renormalized mass) increases as the velocity of the
mass increases. }

\textrm{(4)} \textrm{Summing up the above, Lorentz concluded that the ratio of
the electron's mass in the moving frame and that of the ether frame is }$%
k^{3}\epsilon $\textrm{\ parallel to the direction of motion (longitudinal),
and }$k\varepsilon $\textrm{\ perpendicular to the direction of motion
(transverse) where }$k=\sqrt{1-v^{2}/c^{2}}$\textrm{\ and }$\epsilon $\textrm{%
\ is a constant. Setting }$\epsilon =1,$ \textrm{Lorentz calculated the
expressions for the electromagnetic masses in these directions as}

\[
m_{L}=m_{0}/\sqrt{1-v^{2}/c^{2}}^{3},\quad m_{T}=m_{0}/\sqrt{1-v^{2}/c^{2}}%
\quad \mathrm{where\ }m_{0}\mathrm{\ }=(4/3)(E_{em}/c^{2}). 
\]%
\textrm{This means that well before Einstein's relativity theory, Thomson
and Lorentz concluded that the ``speed of a charged mass" could not exceed }$c$%
\textrm{.\emph{\ }}

\textrm{Indeed,} \textrm{what is astounding here is that this work on
renormalized mass put electromagnetic theory in direct conflict with
classical dynamics. In the latter, mass will not be affected by its speed.
The direct link between speed and mass immediately violates the second law
of Newton where force is determined only by acceleration as mass is
invariant. This manifests itself further as the violation of the second law of
Newton as in the so-called Lorentz force we have }$\mathbf{F}=q(\mathbf{%
E+v\times }\ \mathbf{B}).$

\textrm{All of this implies that the issue of ``electromagnetic mass" is
still alive and to be investigated. The most urgent problem is that
Lorentz force makes the electromagnetic field theory inconsistent as it also uses
Newton's second law. }

\textrm{Regarding Poincar\`{e}'s electromagnetic mass, what is outstanding is
that he obtained it from the electromagnetic momentum instead of the  electromagnetic
energy. J. J. Thomson obtained it from electromagnetic energy. We know that energy is
a modal concept and not a physical reality. }

\subsection{Michelson-Morley experiment }

\textrm{\textsl{As proponents of Newton's dynamics}}\textsl{,} \textrm{%
Michelson and Morley felt urged to find out the absolute speed of our planet in
the absolute space. Their idea was that when we measure the speed of
light in all directions possible, then in the direction of the motion of our
planet the speed of light will be maximum. To their astonishment they found
out that the measured speed of light is the same in all directions. They subsequently concluded that the speed of light is constant }$c$\textrm{\
regardless of the speed of the emitter of light, in symbols }$c+v=c$\textrm{%
. Regarding this apparent contradiction, there are at least two major issues
to be discussed before coming up with hasty conclusions. }

\textrm{(1)} \textrm{Experiment in general is not as simple a matter as
we tend to think. Indeed, the great mathematician/philosopher Russell already
warned us that experimental verification is viciously circular
as to design an experiment, we use a theory whose validity is to be verified by the experiment. In the same sense, he pointed out that experimental refutation of a theory is also viciously circular. To design an
experiment to refute a theory requires the use of the theory to be refuted. This gives
rise to the questioning on the most fundamental role of experiments in
physics, which, unlike engineering, is a theory. Logical coherence is a
most important issue of physics. }

\textrm{(2)} \textrm{It is not quite clear what Michelson and Morley really did. Were
they measuring the speed of light in the absolute frame or in the frame of
the apparatus? All we can say at this stage is that, as \textsl{Newton
implied, the absolute speed of anything cannot be empirically known unless
we stand on the shoulder of God.} Was this not what made Michelson and Morley want
to measure our absolute speed in the universe? }

All of these are rather philosophical and conceptual questions. One
of the major deficiencies of the current practice of physics is the failure to
consider these conceptually important questions. The Michelson-Morley experiment is one of the
main examples of this deficiency. Despite its overvalued
importance, the more we think the more we get lost.

\subsection{Fitzgerald contraction and electromagnetic Lorentz
transformation}

The first group of physicists who took the result of the Michelson-Morley experiment very
seriously were researchers in the electromagnetic field theory, which was expected as the theory
of light was a part of the EM field theory of Maxwell. Fitzgerald prematurely concluded that a moving body shrinks in the direction of
the motion. Theoretically, there is no such thing as physical bodies as
Newton reduced all of them to point masses in order to make the mathematics work to
form a theory of dynamics. This is to say that a body is not the subject of
theoretical study in dynamics. (In modern day term what Fitzgerald called
body is a massively complex system of particles, each of which obeys the
laws of quantum mechanics.) It was Lorentz who embraced this idea of Fitzgerald and developed it into the concept of what we now call
Lorentz transformation. For Lorentz, who was not a relativist, this
transformation however was limited only to between the absolute electromagnetic field and a frame moving inside it. 

\section{Special theory of relativity: kinematics}

\subsection{Special theory of relativity: kinematics}

\textrm{Einstein thought that Lorentz transformation, which was limited just to the
absolute electromagnetic field frame and a moving frame, can be generalized to any two reference frames, thus removing the concept of the absolute frame. As
discussed above, after Newton's dynamics, relativists removed accelerating
reference frames to avoid violating the principle of relativity, the
action-reaction law in particular. Einstein was not an exception. Now the
Galilean relativity theory is limited only to inertial reference frames. }

\subsection{Time dilation and length contraction}

\textrm{To this Galilean relativity theory of inertial frames, Einstein
added the axiom of the Constancy of the Speed of Light (CSL) reflecting the
Michelson-Morley experiment. The CSL says that in all inertial frames the speed of light is
constant }$c$\textrm{. So, if light in a frame F is observed by an
observer A in an inertial frame F and by another observer A' in an inertial frame F', both
A and A' will observe that the same light moves with speed }$c$\textrm{. }

\begin{remark}
\textrm{One of the major issues regarding this claim is that we do not know
what light is and therefore we don't know either what we mean by ``measuring" the speed of light. This
is a near fatal problem with the philosophy of empiricism, which was  questioned by Newton. The concept of measurement is not a definable concept and, as
Russell said, such a concept may well be viciously circular. }
\end{remark}

\textrm{Notwithstanding, based upon the axiom of CSL, Einstein went on to obtain
time dilation (TD) and length contraction (LC) through a thought experiment. Now, the simplest way to explicate Einstein's argument for TD
and LC is to present a thought experiment that exposes the fundamental
inconsistency of the theory of inertial reference frames with the CSL axiom. We
call this thought experiment ``the (power pole)-(power line) paradox.}

\textrm{It goes like this. Assume a train runs on a track. When the tip of the power pole of
the train touches the power line at point P a spark occurs at P. An observer
located in the train straight down the point P will observe that the light
comes straight down to him/her from the point P, which is the tip of the power
pole. Also the observer will see that the same light comes to him/her diagonally from
the point P, which also is a stationary point of the power line. } 
This is a contradiction. and it is consistent with Aristotle's
warning that a point on a line may not be a part of the line. the same is held by contemporary
topologists but in a more modern way, namely that real
numbers (even rational numbers on the real line) are defined through a limit.
Therefore there is no finite access to any real number on a real line. If we
cannot access it, how can we move it? If we cannot move even a single
point in our 3D space, how can we move the entire 3D space inside another 3D
space? In short, topology says that a point does not exist on a
topological space.

\textrm{So what about the issue of TD and LC? The simplest way to refute
these two claims is such that topology refutes the possibility of moving 3D space
inside another space. So, the setting of the thought experiment for proving TD and
LC is invalid. It is no wonder that these two concepts caused all kinds of
contradictions. }

We tend to take mathematics we use in physics lightly just as a
language. This is a perfect example of the price we pay for our ignorance
and arrogance. Mathematical results at the level of topology, etc., are
obtained with utmost care and precision. So often, unless we pay due
attention and effort to understand, we take the results wrongly and end up
with this kind of devastating mistakes.

\subsection{Relativistic Lorentz transformation}

\textrm{Without knowing of the falsity of TD and LC, Einstein showed that
from TD and LC alone, without relating to electromagnetic field theory, we
can obtain Lorentz transformation. From time dilation }%
\[
t^{\prime }=t/\sqrt{1-(v/c)^{2}}. 
\]%
\textrm{and length contraction}%
\[
v^{\prime }=\sqrt{1-(v/c)^{2}}x 
\]%
\textrm{Lorentz transformation is obtained as }%
\[
x^{\prime }=(x-vt)/\sqrt{1-(v/c)^{2}},\quad y^{\prime }=y,\quad z^{\prime
}=z^{\prime },\quad t^{\prime }=(t-vx/c^{2})/\sqrt{1-(v/c)^{2}}. 
\]

\textrm{The proof goes as follows: By applying the effect of length
contraction on Galilean transformation we obtain }$x^{\prime }=(x-vt)/%
\sqrt{1-(v/c)^{2}}.$\textrm{\ Length contraction in the opposite direction is }%
\[
x=(x^{\prime }+vt^{\prime })/\sqrt{1-(v/c)^{2}}. 
\]%
\textrm{\ Solving these two equations for }$t^{\prime }$\textrm{, we have }%
\[
t^{\prime }=(t-vx/c^{2})/\sqrt{1-(v/c)^{2}}. 
\]

\textrm{A common argument for proving TD from LT }%
\[
t^{\prime }=\left( t-vx/c^{2}\right) /\sqrt{1-(v/c)^{2}} 
\]%
\textrm{goes as follows:} \textrm{Set }$x=0$\textrm{, then we have }%
\[
t^{\prime }=t/\sqrt{1-(v/c)^{2}}. 
\]%
\textrm{A more careful logical analysis shows that what this really
showed was that transformed time depends upon the location of the clock! 
\textsl{It did not prove that LT implies TD. To the contrary, it refuted this
claim. To be precise, it showed that when observed at }}$x=0$\textrm{\textsl{%
, time dilates with the gamma factor. TD states that, observed from anywhere
on the} }$x$-\textrm{\textsl{axis, time dilates with the gamma factor. }This
is an interesting instance of the same formula meaning entirely different
things depending upon the context it was obtained in. This is possible because
there is more going on in physics that is visible on the surface of mathematical symbols. }

\textrm{This puts us in a delicate situation where we have to question the
equivalence between Minkowski's special theory of relativity, which does not
use TD and LC, and Einstein's special theory of relativity, which uses them.
This further makes us wonder about the validity of the currently held belief that the general
theory of relativity is a generalization of Einstein's special theory of
relativity. The general theory of relativity includes not Einstein's special
theory of relativity but Minkowski's special theory (tangentially). }

\subsection{Lorentz transformation v.s. principle of relativity}

The Lorentz transformation plays yet other questionable roles. We
can shown that this transformation fails to respect Newton's law of gravitation,
Coulombs' law, Newton's second law and wave equations. For example, despite
the claimed advantage of conserving wave equations, Lorentz transformation
astoundingly fails to conserve the more fundamental second law and the law
of gravitation.
\textrm{Considering the way Lorentz transformation was obtained, it is
not surprising that these two major laws of mechanics are not Lorentz
invariant. This means that \textsl{Lorentz transformation is not a relativistic transformation as it violates the principle of relativity}. 
}

\subsection{Is wave equation invariant under the Lorentz
transformation?}

\textrm{We can further show that the claimed invariance of wave equations
under Lorentz transformation is false. To make the argument more articulate,
let us discuss the issue under a general setting. }

\begin{eqnarray*}
\frac{\partial \psi (x^{\prime },t^{\prime })}{\partial x} &=&\frac{\partial
\psi (x^{\prime },t^{\prime })}{\partial x^{\prime }}\frac{\partial
x^{\prime }}{\partial x}+\frac{\partial \psi (x^{\prime },t^{\prime })}{%
\partial x^{\prime }}\frac{\partial t^{\prime }}{\partial x} \\
&=&\frac{\partial \psi (x^{\prime },t^{\prime })}{\partial x^{\prime }}\frac{%
\partial \gamma (x^{{}}-vt)}{\partial x}+\frac{\partial \psi (x^{\prime
},t^{\prime })}{\partial x^{\prime }}\frac{\partial \gamma (t-\frac{vx}{c^{2}%
})}{\partial x} \\
&=&\gamma \frac{\partial \psi (x^{\prime },t^{\prime })}{\partial x^{\prime }%
}-\frac{\gamma v}{c^{2}}\frac{\partial \psi (x^{\prime },t^{\prime })}{%
\partial t^{\prime }}
\end{eqnarray*}%
\textrm{Similarly}

\[
\frac{\partial \psi (x^{\prime },t^{\prime })}{\partial t}=-\gamma v \frac{%
\partial \psi (x^{\prime },t^{\prime })}{\partial x^{\prime }}+\gamma  \frac{%
\partial \psi (x^{\prime },t^{\prime })}{\partial t^{\prime }} 
\]%
\[
\frac{\partial \psi ^{2}(x^{\prime },t^{\prime })}{\partial x^{2}}=\left(
\gamma \frac{\partial }{\partial x^{\prime }}-\frac{\gamma v}{c^{2}}\frac{%
\partial }{\partial t^{\prime }}\right) \left( \gamma \frac{\partial }{%
\partial x^{\prime }}-\frac{\gamma v}{c^{2}}\frac{\partial }{\partial
t^{\prime }}\right) =\gamma ^{2}\frac{\partial ^{2}}{\partial x^{\prime 2}}-2%
\frac{\gamma ^{2}v}{c^{2}}\frac{\partial ^{2}}{\partial x^{\prime }\partial
t^{\prime }}+\frac{\gamma ^{2}v^{2}}{c^{4}}\frac{\partial ^{2}}{\partial
t^{\prime 2}} 
\]%
\textrm{Similarly }%
\[
\frac{\partial \psi ^{2}(x^{\prime },t^{\prime })}{\partial t^{2}}=\gamma
^{2}v^{2}\frac{\partial ^{2}}{\partial x^{\prime 2}}-2\gamma ^{2}v^{{}}\frac{%
\partial ^{2}}{\partial x^{\prime }\partial t^{\prime }}+\gamma ^{2}\frac{%
\partial ^{2}}{\partial t^{\prime 2}} 
\]%
\textrm{This is valid only under the condition }$v=c=\omega .$\textrm{\ The
second equality comes from the fact that }$\omega $\textrm{\ is the wave speed. The
first equation implies that the frame speed is }$c$\textrm{\ which is not
possible in the special theory of relativity. This means that Einstein's claim
that the electromagnetic wave equation is invariant under the Lorentz
transformation is invalid. It is a well understood fact that there is no
reference frame for light at the pain of contradiction. If }$v=\omega $ 
\textrm{then the gamma factor becomes undefined and there is no Lorentz
transformation for such a frame. }

\textrm{All of this was well expected logically. Lorentz transformation is
defined in terms of the constant }$c$\textrm{, which is the speed of
electromagnetic waves in vacuum. So, there is no convincing reason why this
transformation will conserve wave equations, which are not electromagnetic
wave equations of Maxwell. }

\subsection{Inconsistency of the special theory of relativity}

\subsubsection{The power pole - power line paradox}

\textrm{Assume a train runs and when the tip of the power pole of the train
touches the power line at point A, spark occurs at this point. An observer
in the train located straight down the tip of the power pole will observe
that the light comes straight down from the tip of the power pole. However
as the point A is also a stationary point of the power line, he will observe
that the light comes diagonally down from the stationary point A. This is a
contradiction. }

\textrm{This contradiction tells us that we cannot move one reference
frame, which is a 3D space, inside the other. Aristotle knew this 3,000 years
ago and said that a point on a line will not be a part of the line. Modern
topologists will say that a point on a real line is accessible only through
infinite limit process and so we have no finite access to any point on a
real line. All of this means that we cannot move a point in a space. Then
how is it possible to move an entire 3D space inside another. This paradox
tells us that even the Galilean theory of relativity is inconsistent. }

\subsubsection{Deductive paradox}

\textrm{The general theory of relativity (GTR) deduces }$c+v=c+v.$\textrm{\ Einstein added the CSL axiom }$%
c+v=c $ \textrm{to GTR\ to form the special theory of relativity (STR) kinematics. The outcome of this is that the
STR kinematics proves both CSL and its negation. This is a deductive
inconsistency. The problem here is that by adding a new axiom to an old theory,
one cannot block any theorem deducible from the old theory. It is called
the monotonicity of deduction and is a most basic law of formal reasoning. }

\subsubsection{Speed paradox}

\textrm{Assume A observes that B is moving with constant speed }$v=d/t$%
\textrm{\ relative to A (and to each other), where }$d$\textrm{\ and }$t$\textrm{\ are
length and time observed classically by A. Then B will observe that A is
moving with speed }$v\prime =d\prime /t\prime $\textrm{\ where }%
\[
d\prime =d/\sqrt{1-(v/c)^{2}},\qquad t^{\prime }=t\sqrt{1-(v/c)^{2}}. 
\]%
\textrm{\ Then we have }$v\neq v^{\prime }.$ \textrm{The problem here is
that [speed] is determined by the [length] and [time] \ as }%
\[
\lbrack speed]=[length]/[time]. 
\]%
\textrm{So, [speed] cannot alter [length] and [time] as in length
contraction and time dilation. This is to say that STR kinematics violates
a most fundamental law of dimensional analysis. }

\textrm{This can be argued in a more ontological way as follows: Assume an
object O in another frame moves inside our frame a distance }$d$\textrm{\ in
time }$t$\textrm{. Let A and B be the positions of O at time }$0$\textrm{\
and }$t$\textrm{\ in out frame, respectively. According to the perspective of
O, A and B are moving points. So, O will observe that the distance }$%
d^{\prime }$\textrm{\ between A and B is less than }$d$\textrm{. Also O will
observe that the time interval }$t^{\prime }$\textrm{, which took A to be
in front of it and B to be in front of it, is more than }$t$\textrm{. So, O
will observe that the speed of our frame is }$d^{\prime }/t^{\prime }$%
\textrm{, which is not }$d/t$\textrm{. This failure means that according to
the relativity theory an observer observes the speed of a reference frame as }%
$v$\textrm{\ and as }$v\prime $\textrm{\ such that }$v\neq v^{\prime }.$

\subsubsection{Dingle's paradox}

\textrm{Herbert Dingle knew STR kinematics extremely well. He was one of the top
researchers of STR before he became a top critic of this theory. He knew
well that one cannot consider acceleration in this theory because of the
principle of relativity. He thus considered two clocks that are already
synchronized and moving towards each other with constant speed. He pointed
out that due to the time dilation each clock will see the other moving slower. }

\textrm{Main stream relativists responded arguing that when one twin moves
out and comes back, then it is this twin who went through time dilation and
not the other twin. So, there is no paradox, according to this argument. 
However, we
are discussing time dilation that is to occur in STR kinematics. So, the
element of acceleration is out of the issue. It might be that STR dynamics
may support this argument. But certainly such theory, if any, should bring
back the same problem for the symmetric acceleration case. If both twins
accelerate symmetrically and come back to see each other, we have the same
problem. This means that the claimed generalization of STR kinematics to
STR dynamics also has a serious inconsistency problem. }

\subsection{Michelson-Morley experiment revisited I}

\textrm{There are two criticisms of the way how this historic experiment was
treated in theoretical physics. This experiment was interpreted
under the assumption that light is an electromagnetic wave, in accordance with Hertz's proposal.
Even to this day though, we are not sure what light is. We were told to
accept the view of Hertz without a substantial proof thereof. }

\textrm{Moreover, electromagnetic waves are not physical reality as the
concept of electromagnetic field is not a physical reality. This concept is
what logicians and philosophers call ``modality", ``counter factual modality"
to be precise. The spatial distribution of electromagnetic force per a unit
charge is not reality. Such distribution should appear in reality only when
we place unit charge everywhere in the space. But, if we place a unit
charge at every point in the electromagnetic field, the source which created
such a field will be affected and the electromagnetic field will not be
maintained. Moreover, the placed unit charges will react with each other making
it impossible to sustain such a configuration. \textsl{Putting it more mathematically,
there is not enough charges to fill all points in a space. Pure
mathematics will put this as follows: There are uncountably many points
in a geometric space and there are only countably many charges in the
universe.} This implies that the concept of electromagnetic field has nothing to do with reality. To be precise, the so-called electromagnetic
wave is an ``action at a distance transmission" of the change in
electromagnetism at the source to a charge placed at a certain location in
the space. There is no wave. This is precisely why without wave medium the
so-called electromagnetic waves ``travel" with speed }$c$\textrm{. So, there
is no need for the ``aether". There is no physical realism that 
supports the counter-factual modality. All of this calls to question 
the treatment of light as an electromagnetic wave by the Michelson-Morley experiment. }

\section{Special theory of relativity: dynamics}

\subsection{Einstein's ambition and its fallout}

Galilean theory of relativity did not consider reference frames
that are under acceleration relative to each other. This was because
acceleration, through the second law, violates the principle of relativity.
This only restriction imposed by kinematics on relativity theory was too limiting for
Einstein. Considering that way before this setback, already at the most
basic level of Galilean theory of relativity the concept of relativity is
insurrectionist, Einstein should have abandoned the idea of relativity. It is unfortunately not what happened.

\subsubsection{Relativistic collision, relativistic mass,
relativistic momentum and relativistic energy}

\textrm{Einstein's first motive towards STR dynamics was to consider the 
relativistic collision problem. To make sure that the conservation of momentum
holds for relativistic collision, Einstein defined relativistic mass as follows:}%
\[
m=m_{0}/\sqrt{1-(v/c)^{2}} 
\]%
\textrm{where }$m_{0}$\textrm{\ is the rest mass and }$v$\textrm{\ is the
speed of the mass in relation to the observer.} \textrm{From this he obtained the famous
relativistic energy formula as follows: The relativistic second law is }%
\[
\mathbf{F}=d\mathbf{p}/dt 
\]%
\textrm{where }$\mathbf{p}=m\mathbf{v}$\textrm{\ is the relativistic
momentum.} \textrm{\ Then }%
\[
dE=\mathbf{F}\cdot d\mathbf{r=}\frac{d(m\mathbf{v})}{dt}\cdot d\mathbf{r}=d(m%
\mathbf{v})\cdot \mathbf{v=}dm\mathbf{(v\cdot v)+}m(d\mathbf{v}\cdot \mathbf{%
v}). 
\]%
\textrm{From this he calculated that} 
\[
E=mc^{2} 
\]%
\textrm{The mistake made here is that Einstein forgot that up to here }$\mathbf{v}
$\textrm{\ is constant. For mathematical sanity, in collision problems, we do
not consider accelerated bodies. The moment of impact is excluded from 
consideration. So, what we should have here is }%
\[
E=0 
\]%
\textrm{rather. This mistake impacted heavily the relativistic energy-momentum relation}

\[
E^{2}-c^{2}p^{2}=m_{0}^{2}c^{4}. 
\]%
\textrm{However, from the point of view of dimensional analysis, this is
not surprising at all. There is no reason to think that the dimension of
energy and that of momentum are related. \textsl{Indeed, energy is not even
a physical dimension. It is a modality rather as it is the ``potential" to do
[work]}. So, [work] is a physical dimension but energy is not. Therefore \textsl{the energy-momentum relation of Einstein is false
conceptually as well.} }

\begin{remark}
\textrm{Indeed, there is a complaint from the discipline of wave mechanics regarding
Einstein's energy-momentum relation. In wave mechanics of continuum medium,
there is nothing moving in the direction of the motion of the wave. The only
thing that is moving in this direction is the local vibration of the
medium. This means that there is no momentum in waves. }
\end{remark}

\subsubsection{Impact on quantum field theory}

\textrm{All of this implies the end of the entire 20th century
theoretical physics. Einstein correctly said that when his relativistic
energy equation }$e=mc^{2}$ \textrm{fails, the entire 20th century
theoretical physics fails. }

\textrm{Indeed,} \textrm{upon this energy equation are based both the duality of the photon and the electromagnetic wave }

\[
E=h\nu =pc\qquad h=h/\lambda 
\]%
\textrm{as well as the de Broglie relation, paving way to what is called
quantum mechanics. Also, as we will discuss later, Gordon-Klein's
theory of relativising quantum mechanics by replacing }$E$\textrm{\ and }$p$%
\textrm{\ with quantum operators \textbf{E} and \textbf{p} in the energy-momentum relation is invalid. The entire quantum field theory also
collapses. The same convention used by Dirac in his quantum electrodynamics also is invalid for
the same reason. }

\subsubsection{More contradictions coming from $e=mc^{2}$}

\textrm{The quantum mechanics, which was built upon the special theory of
relativity, quantized light as electromagnetic wave and presented what we now call\
``photon" as the particle dual of light wave. To avoid the famous
relativistic formula}%
\[
e=mc^{2}=m_{0}c^{2}/\sqrt{1-v^{2}/c^{2}} 
\]%
\textrm{diverging for the photon with }$v=c$\textrm{, Einstein assumed that for
the photon the rest mass }$m_{0}=0.$\textrm{\ This lead to }$e=0/0$\textrm{\
which Einstein thought can be any number as the linear equation }$0x=0$%
\textrm{\ has any number as its solution. This is wrong because }$0x=0$\textrm{\
does not involve the division by }$0$\textrm{\ while }$e=0/0$\textrm{\
involves division by }$0$\textrm{, which is an impossible operation. This rather
expectedly leads to the following contradiction:}%
\[
E=\sqrt{(cp)^{2}+(m_{0})^{2}c^{4}}=cp=m_{0}vc/\sqrt{1-vc^{2}}\frac{{}}{{}}%
=(0/0)cv=c^{2}h\nu =h\nu . 
\]%
\textrm{We can derive yet another contradiction: }%
\[
E=\sqrt{(pv)^{2}}=\sqrt{c^{2}m_{0}/\sqrt{1-(v/c)^{2}}=\sqrt{0/0}=\sqrt{h\nu }%
}=h\nu =1. 
\]%
\textrm{Without knowing this problem, photons are now introduced as a
legitimate particle dual to light wave with rest mass }$0$\textrm{\ and
speed }$c$\textrm{. What is truly paradoxical is that a particle that never
rests now has a rest mass }$0$\textrm{. This is what philosophers and
logicians call a category error. }

\subsection{Michelson-Morley experiment revisited II}

\subsubsection{Light-as-photon interpretation of Michelson-Morley experiment}

\textrm{With the photon becoming a particle dual to the light wave, we have yet another interpretation of Michelson-Morley experiment. ``%
\emph{Assume photons are particles."} Then it must be the case that when we
emit a photon to the vacuum from an emitter moving with speed }$v$%
\textrm{, the speed of the photon must be }$c+v$\textrm{\ in the vacuum. So,
the emitted photon moves towards the reflecting mirror with speed }$v+c$%
\textrm{. But as the mirror itself moves with the speed }$v$\textrm{\ in the
same direction, the effect of }$v$\textrm{\ cancels. When the photon is
reflected at the mirror, it comes out with speed }$c-v$\textrm{. But as the
receiver of this photon is moving towards the photon with speed }$v$\textrm{, this }$v$\textrm{\ cancels again. This means that this experiment will
not detect the }$v$\textrm{. In conclusion, }$c+v=c+v$ \textrm{but the
experiment set as it is cannot detect this }$v$.

\subsubsection{Quantum mechanical interpretation of Michelson-Morley experiment}

\textrm{Experimental physics sometimes asserts that in order to
understand the Michelson-Morley experiment, one must consider the quantum mechanical process
of light reflecting at the mirror surface. If reflection of light at the
surface of a mirror is the consequence of energized (by the incident light)
electrons inside the mirror surface recoiling, then it should take some time
for the reflected light to come out of the surface of the mirror.
Considering the Compton effect, it may not be the case that the reflected
light is not of the same frequency as the incident light. It is not clear
how seriously these questions were taken in theoretical physics. }

\textrm{Recently, Wheeler carried out an experiment using an apparatus
 very similar to the Michelson-Morley apparatus which however used a
half-silvered mirror instead. The experiment showed an unexpected behaviour of this
apparatus which current quantum mechanics cannot explain. The problem with
the so-called ``splitter" (half-silvered mirror) is that it had an
explanation only for a pure wave theory of light, but it has no
explanation using the light quanta. Better said: if photons are deviated
with probability 50 \%, there is no interference. But on the other side, if
they are really split in two directions, this contradicts the Planck-Einstein
hypothesis, as far as the two pieces of a photon are entangled and are
evidently waiting for the two halves to build a photon together. So in a certain way,
the interference does not indicate anymore some delay of a half-ray
relatively to the other, as planned by Michelson-Morley. This makes us wonder
if Michelson-Morley's interpretation was correct. }

\textrm{According to the model used by Michelson-Morley, in the second
splitter (or by twice passing the same splitter) one should have had again a
50 - 50 distribution, and not a 100 - 0 distribution, as observed by
Wheeler. Without intending it, Wheeler shows with his experiment that the
Michelson-Morley apparatus contains aspects that nobody has thought about. So one
cannot claim to have experimental proof of the constancy of the speed \ of light
using something that nobody understands.}

\textrm{This problem exhibits exactly the same pattern as the Michelson-Morley
experiment, which was conducted under the assumption that light was a wave.
The irony is that, using Michelson-Morley experiment, through the special theory
of relativity the ``wave-particle duality" was introduced first and then the ``particle
theory of light" offered entirely different picture. This way,}\textsl{\ 
\textrm{in the end, the empiricism proved that the wave-particle duality
hypothesis is invalid at the pain of contradiction. }}

\subsection{From Einstein, through Dirac to material science: the particle-wave duality in full swing}

\textrm{The ultimate product of this highly questionable wave-particle
duality manifested itself most dramatically in the quantum electrodynamics (QED) of
Dirac. Quantum electrodynamics seems to be the ultimate end-product of
Einstein's special theory of relativity, which is inconsistent. Here is a
summary of the complications we have regarding this issue. }

\begin{enumerate}
\item \textrm{\textsl{The reflection of light on the mirror is a complex
problem of statistical particle physics and only material science can bring
us closer to the truth. The problem we have here is that we have not such
material science. Also the material science which digs into this kind of
problems must come from a satisfactory quantum mechanics, which we have not,
as our quantum mechanics is based upon the special theory of relativity, which
came from the Michelson-Morley experiment. To make the matter even worse, now
Wheeler's result seems to support the concern that Michelson-Morley's experiment could have been
interpreted wrongly. }}

\item \textrm{\textsl{Even if the \textrm{Michelson-Morley experiment} is
limited to electromagnetic waves only, these waves are not physical reality. They are counter-factual modality. Moreover, when we operate upon the assumption of wave-particle duality,
the uncertainty principle creeps in and this makes the constancy of the speed of
light claim ``statistical". When a most fundamental assumption of our theory
is of statistical nature, we do have some serious concern. True, the
original CSL argument is not statistical. It was
based upon abstract (not physical) wave theory of counter-factual modal
waves. But the recent argument by Wheeler again is statistical. }}
\end{enumerate}

\textrm{Early quantum mechanics in principle avoided getting into this kind
of problems when the problem of particles enclosed by walls etc. was considered. Walls are represented not as a complex material, but
as potential barriers, which is nothing but a mathematical entity. Here,
one is trying to study the reflection of light at the mirror. Quantum
mechanics cannot really handle the reflection of light on a mirror as we do
not understand what mirrors are sufficiently enough to discuss these issues. To
understand it requires perfect quantum mechanics, which we have not. }

\textrm{We use macro materials, which are microscopically incredibly complex,
to do our experiments. It is not a problem when we do macro-level physics. But
when we deal with micro-level physics, we have no solution. Our experimental
instruments belong to the macro-level physics. The cosmology shares the same
problem. We can never conduct experiment or measurement at this incredibly large
scale level. It is not promising to do physics of the cosmos through looking
at billions of stars light years away from us. The only tool we have available here is
the relativistic Doppler effect and we do not even know what light is. }

\textrm{What is striking is that regardless of the status of quantum
mechanics, Wheeler's experiment shows that while passing the second splitter,
the light ``knows" which part it came out from the first splitter. In the
second splitter we get the Hadamard-Walsh gate, which is an ``operator" in
quantum computing. This is sufficient for engineering. But it is unfortunate that
our theory will not make us understand all of this. }

\textrm{Philosophically, what we are facing is a crisis where the most basic
laws of physics are no longer macroscopic laws. To make it macroscopic we went through a statistical argument, which makes us wonder what laws of
physics are. }

\textrm{Moreover, all of these problems by which we become overwhelmed make us wonder what do we really mean by experimental observation, which is
the essence of empiricism. To make experiments at the micro-level, we need a
functional theory of the micro-level physics. But such a theory must come from an 
acceptable micro-level experiment. We are going around a big circle. As we
said above, cosmology is facing the same difficulty at the other end of the
spectrum. In the case of micro-level physics, we have a distinguished difficulty
of uncertainty issue which makes the theory probabilistic. All of these
issues have to be dealt with even after we manage to free quantum physics
from the inconsistency of relativity theory. A long way to go lies ahead of us. }

\textrm{So, it is no longer just an isolated problem of theoretical physics.
The dynamic linkage between theory and empiricism has to be re-examined and
new working link must be established. 
}

\section{Minkowski's relativity theory}

\textrm{It appears to be the case that Minkowski's 4D spacetime relativity theory was
a serious effort to make the inconsistency problems in Einstein's special
theory of relativity disappear mathematically. }

\textrm{For Einstein, the Lorentz transformations are transformations from
one inertial 3D frame that is moving inside another 3D frame with
relative speed }$v$\textrm{\ and the associated transformation of time. This
immediately lead to the flooding of contradictions, which we discussed in the
foregoing. Most of such paradoxes originate from the ``power line-power pole
paradox" which is deeply embedded in the theory of Galilean inertial
reference frames upon which Einstein's STR was built. As a logical
triviality, \textsl{inconsistency problems will not disappear by adding one
extra axiom of CSL.} }

\textrm{The reason behind Minkowski's apparent success is that his theory
appeared to have little to do with the troubled part of Einstein's STR,
which is based upon mutually moving 3D reference frames. From this troubled
assumption Einstein proved TD and LC, which naturally lead to a mountain of
paradoxes. Lorentz and Einstein in their own setting deduced LT from TD and
LC. Unfortunately, motion takes place inside a geometric space and so we cannot move reference frames, which are essential for defining such motion. 
\textrm{This issue was already philosophically addressed by Aristotle. He
pointed out that a point in a geometric space is not a part of the
space. Modern topology explicated this warning of Aristotle by proclaiming that there is no point in a geometric space.}
it said rightly that a point in a geometric space can be ``accessed" only
through limit. \textsl{If we cannot reach it by finite means, how can we
``move" it?} This kind of conceptual issues are quite well attended to in pure
mathematics as mathematics has experienced the horror of contradictions at
deep levels. Also, the lack of empiricism made mathematicians focus on these
conceptual problems. }

\textrm{Here is a possible interpretation of Minkowski's work as a mathematical
physicist. He used only one 4D spacetime as a reference frame (plus time)
and from his single reference frame, he ``\textsl{not formally but
conceptually}" derived two reference frames, say F1 and F2. First, he placed
F1 as the ``mother frame" and defined a Lorentz transformation from it to itself where }$v$\textrm{\
in the gamma factor is the mutual speed between F1 and F2. In this way he
thought he ``\textsl{simulated}" F1 and F2 with just F and the Lorentz transformation from F to
itself. Unfortunately, his formalism by itself did not explicitly support
this interpretation. This is a possible but rather obscure interpretation
of his formalism.}

\textrm{Minkowskian relativity theory with Minkowski distance dominated the
entire theoretical physics for nearly a century as the deepest theoretical
foundation of physics that apparently evaded the contradiction associated
with Einstein's special theory of relativity. When under the help from Hilbert, Einstein accepted
Minkowskian spacetime as a local tangential spacetime at each location in
the Riemannian spacetime theory of the general theory of relativity, the
``consistency" of STR and GTR was ``established" and relativity theory became
the ultimate truth in theoretical physics. }

\textrm{What is interesting here however is that this Minkowskian 4D
spacetime approach can be adopted to formulate the issue of Fitzgerald
contraction (LC) by considering the unique 4D spacetime as the universe and
the Lorentz transformation as a representation of a specific observer frame and }$v$\textrm{\
representing the speed of the observer frame inside the universal frame. So,
Minkowski's theory integrates the issue of relativity in the setting of
electromagnetic field theory. However, all we could do for this new
formulation of Einsteinian STR was to hope that the equivalence of Lorentz transformation and
(TD,LC) will hold.\textsl{\ It did not happen, as we have shown in the foregoing. }}

\textrm{After all, all of this is a pointless discussion as (TD, LC) pair
deduce ``physical paradoxes". Deducing (TD, LC) from Lorentz transformation simply removes the 
apparent capacity of Minkowski's theory to be a consistent alternative to \
Einstein's relativity theory. Moreover, even if Lorentz transformation did not deduce (TD, LC)
and saved itself from inconsistency, we would have ended up with the question of the
relevance of such a theory in theoretical physics. Under the standard
interpretation of inertial reference frames, as we have discussed, LC
implies TD. }

\textrm{As Einstein's original SRT is plagued by the inconsistency, the
uncertainty of the status of Lorentz transformation in Minkowski's theory seems to be the only
hope left. The apparent discrepancy between Lorentz transformation and (TD, LC) is still
giving us some hope. }

\textrm{The problem here is that we do not know what Lorentz transformation means physically if }%
$v$\textrm{\ is not the relative speed of two reference frames and Minkowski did
not define Lorentz transformation using two reference frames. Nevertheless, there is an
understandable reason why he did not use two 4D spacetime frames. If we use
two then there is not much point in using 4D spacetime. Einstein's theory is
easier to handle. In Einstein's theory, 3D space and time are independent
and separate. So, technically it is easy to consider a 3D space move inside
another space and vice versa. The only problem with this is that this leads to
the geometric paradox which killed Einstein's STR. But there is yet another
question here. It is mathematically impossible to discuss the motion of 3D
space inside the 4D spacetime. All we can do is to express motions in the 3D
space as geodesics inside the 4D spacetime. But is this not all we need? }

\textrm{We have a mountain of nontechnical, extremely challenging and deep
problems on the border of physics, mathematics and philosophy. And literally,
nothing has been done. }

\textrm{One of the most important contributions of Minkowski was the
metric on his 4D spacetime. This came from the mathematical argument that
his metric }$d\tau $ \textrm{such that} 
\[
(d\tau )%
{{}^2}%
=(dt)%
{{}^2}%
-(1/c)((dx)%
{{}^2}%
+(dy)%
{{}^2}%
+(dz)%
{{}^2}%
) 
\]%
\textrm{is invariant under the Lorentz transformation. It was understandable that Minkowski had
to look for such a metric as Lorentz transformation changes the metric on [time] and that on
[length] relative to the [speed] }$v$\textrm{. Luckily he found one. It is
neither a space metric nor a time metric as Lorentz transformation operates upon 4D spacetime. Hence, there
are some issues:}

\begin{enumerate}
\item \textrm{This metric does not form a topological metric space over the
4D spacetime and this is totally expected as Minkowski adopted the
irregularity inherent in relativity theory that [speed] = [length]/[time]
redefines [time] and [length], which leads to contradiction. That Lorentz transformation 
conserves this metric is yet another indication of the highly questionable
status of this transformation. }

\item \textrm{This metric could be negative, which makes no sense at all.
There is no such thing as a distance that is negative. Regardless of the
direction of an arrow, the length of the arrow is always positive. However, it may be due to the fact that an inconsistent theory can produce any result. This is why such a theory is useless. }

\item \textrm{Minkowski's metric is appreciated as it defined light cone
interpretation. Light cone is formed inside absolute space time. After the
introduction of the light cone, thus, there was a tendency to stop using relativity
theory and just stick to the absolute frame, a very distorted absolute
frame. }
\end{enumerate}

\textrm{It is an open question what this metric means mathematically and philosophically. It means that maybe Minkowski relativity theory is consistent but
with the cost that it has no relevance to anything including physics at all.
After all, Lorentz transformation came from TD + LC, which is inconsistent. There are more
questions than answers. }

\textrm{Given that the Minkowski equation }

\[
(d\tau )%
{{}^2}%
=(dt)%
{{}^2}%
-(1/c)((dx)%
{{}^2}%
+(dy)%
{{}^2}%
+(dz)%
{{}^2}%
) 
\]%
\textrm{is invariant under the Lorentz transformation, is it so also under the
TD and LC? The answer is a ``no". A proof goes as follows: }%
\[
dt^{\prime }=\sqrt{1-v^{2}/c^{2}}dt\quad dx^{\prime }=dx/\sqrt{1-v^{2}/c^{2}}%
\quad dy^{\prime }=dy\quad dz^{\prime }=dz 
\]%
\textrm{Therefore}%
\begin{eqnarray*}
&&(dt^{\prime })^{2}-(1/c)((dx^{\prime })^{2}+(dy^{\prime })^{2}+(dz^{\prime
})) \\
&=&(1-v^{2}/c^{2})(dt)^{2}-(1/c)((dx)^{2}/(1-v^{2}/c^{2})+(dy)%
{{}^2}%
+(dz)%
{{}^2}%
)) \\
&\neq &(dt)%
{{}^2}%
-(1/c)((dx)%
{{}^2}%
+(dy)%
{{}^2}%
+(dz)%
{{}^2}%
).
\end{eqnarray*}%
This reconfirms that Einstein's STR and Minkowski's STR are two
different theories. There is no such thing as Minkowski distance in
Einstein's STR. There is no light cone either. This is a good news in a
sense as the inconsistency of Einstein's STR will not be deleterious to  Minkowski'd STR.
However, as we have stressed many times, nobody knows what Minkowski's STR\ is and
\ what it is for. There is no ontology associated with it. Furthermore
we now have to cleanly detach Minkowski's STR from Einstein's STR. It is a
lot of work, especially because most of popular results in STR came from
Einstein's version. This is however expected, because it is not clear what Minkowski was talking about.

\section{General theory of relativity}

\textrm{Einstein resolved the problem of the limitation of STR\ kinematics
\ by violating the principle of relativity, as in the development of the STR
dynamics leading the entire theoretical physics to the fallacy of }$%
e=mc^{2}. $\textrm{\ Next, he ventured into a new
theory in which accelerating frames can be treated as inertial frames with
gravitational field induced by acceleration. }

\subsection{Principle of equivalence}

\textrm{Einstein assumed that if an accelerating reference frame can be
reduced to an inertial frame in which acceleration induces ``gravitational
field"}, \textrm{then it is possible to treat accelerating frames as inertial
frames within the theory of relativity, which rejects reference frames under acceleration for a legitimate reason. He called this the} \textrm{%
\textbf{principle of equivalence}}\textbf{.} \textrm{To that end he proceed
as follows: \ }

\textrm{Assume a spaceship is in inertial motion in our reference frame.
Moreover, a force accelerates this spaceship with a rate }$\mathbf{\alpha }.$%
\textrm{\ A body }$m$\textrm{\ in the spaceship experiences a force }$%
\mathbf{f}$\textrm{, which is due to the acceleration of the spaceship that makes the body }$m$\textrm{\ move with an acceleration of rate }$\mathbf{a}$%
\textrm{\ in the frame of the spaceship. Putting aside what the force }$%
\mathbf{f}$\textrm{\ is, this means }$\mathbf{f}=m\mathbf{a}.$\textrm{\ Then
from our perspective, }$m$\textrm{\ in the spaceship experiences the
acceleration with rate }$\mathbf{\alpha }+\mathbf{a}.$ \textrm{So, }$m$%
\textrm{\ will experience }$\mathbf{f}=m(\mathbf{\alpha }+\mathbf{a}).$%
\textrm{\ Therefore,}%
\[
\mathbf{f}-m\mathbf{\alpha }=m\mathbf{a}\quad \quad \quad \quad \quad
\quad \quad \textrm{(IF)}
\]%
\textrm{This means that, from our perspective, the acceleration }$\mathbf{%
\alpha }$\textrm{\ on the spaceship induces an ``additional" force }$-m%
\mathbf{\alpha }$\textrm{\ on }$m,$\textrm{\ which Einstein called ``inertial
force" upon the mass }$m$\textrm{\ and the equation (IF) yields the force }$%
m $\textrm{\ experiences in the accelerating spaceship. This Einstein called
the second law in the accelerating frame of the spaceship. According to
him, upon the modification of }$\mathbf{f}$\textrm{\ to }$\mathbf{f}-m%
\mathbf{\alpha }$\textrm{, the second law is conserved under the choice of
accelerating reference frames. }

\textrm{There are several issues to be considered with respect to that.}

\begin{enumerate}
\item \textrm{Putting aside the invalidity of the special theory of
relativity (as we discussed in the foregoing), some issues have been overlooked here. Namely, according to the special theory of relativity, even
addition of speeds is not classical addition. One has to use the so-called
relativistic addition of speeds }$v\oplus v^{\prime }.$\textrm{\ So, how can
the addition of accelerations be the same as classical addition, hoping that in the words of Einstein, acceleration is still the time derivative of speed.
This confusion is closely related to the issue of the disjointedness
between the acceleration and relativistic reference frames. For that reason
the special theory of relativity does not consider acceleration, as we have been told. }

\item \textrm{As an important example, this inertial force is also closely
related to the issue of ``fictitious force" on a mass inside an orbiting
object. Fictitious force is a ``force in fiction", not reality. The reason
why we have a problem with the fictitious force for an orbiting spaceship is
because orbiting spaceships are under centripetal acceleration. Therefore, it is not an
inertial frame. It is wrong to claim that just because inside the orbiting
spaceship a fictitious force called centrifugal force is ``created?", the
orbiting spaceship becomes an inertial reference frame. }

\item \textrm{This fictitious force is the creation of the relativistic
interpretation of the second law of Newton. This law can be interpreted in
two ways. First, when force }$\mathbf{f}$\textrm{\ is applied to a mass }$m$%
\textrm{\ , it accelerates the mass with the rate }$\mathbf{a}=\mathbf{f}/m.$
\textrm{Second, when }$m$\textrm{\ is accelerating with a rate }$\mathbf{a}$%
\textrm{, a ``fictitious force" }$\mathbf{f}$\textrm{\ appears. Ontologically
the second interpretation is invalid. It is always the case that a force
applied to a mass causes an acceleration of the mass. It becomes clear when we consider the
case of an accelerating train. Assume a stationary train
accelerates. Then, a passenger in the train will feel that he/she is pushed back
(against the direction of the train's acceleration). This is simply because
the observer tries to stay where he/she is in the train due to the first law of Newton. If
the observer can see the outside, which is not moving, he/she will see that it is not him/her
but the train that is moving under the force. }

\item \textrm{Now it is clear that the problem of inertial force (fictitious
force) is caused by the faulty relativistic interpretation of the
second law of Newton. This is to say that the second law is not
``relativistic", further confirming that the relativism as per
Galileo and Einstein is untenable. With this, we have that the relativism violates not only the third law of Newton but also the second law. }

\item \textrm{It also is important to note that the fictitious force
violates the third law of Newton. }
\end{enumerate}

\subsection{Violating the point mass assumption}

\textrm{An even} \textrm{more fundamental issue here} \textrm{is considering
the spaceship (or train). In the theory of dynamics, as Newton made it clear,
theoretically there is no such thing as a spaceship. All physical bodies
must be point masses, and rightly so. For this reason, there is no such thing as a mass }%
$m$\textrm{ inside a spaceship. In dynamics, there is no
such thing as a spaceship to begin with, for very good mathematical reasons,
as incisively explained by Newton. }

\begin{remark}
\textrm{Here it is important to notice that there are two kinds of the
violation of point mass assumption in post-Newtonian physics. First, mass taken as a 
solid with volume. This situation can be dealt with ease by reducing the
mass to a point mass, as we are used to. Second, a mass with an inside space
such as a spaceship or a train. This case is a lot more complicated as it may
allow another mass in the inside space. When we collapse the outer mass, what
happens to the inside mass? }
\end{remark}

\textrm{Moreover, a body attached firmly to the wall of a spaceship 
will not experience any special force. This is because when we
consider the dynamics of the spaceship it is just a point mass and there is
no ``inside" of it to which this extra mass is to be attached. Even if we
reluctantly allow spaceships, according to the law of inertia without
force exerted, a mass will continue constant speed motion in a frame. So,
what force is supposed to be exerted upon this body ``inside" the
spaceship"? Does this firmly attached mass move inside the spaceship? }

The usual response is that we experience such force even if we are
firmly attached to the inside wall. 

However, our body is not just a solid. 
\textsl{Our body is beyond the category of physical objects.} Our body has
incredibly complex internal system for perception. This is why we feel such a 
pressure.

\textrm{The most fundamental reason why Newton correctly reduced all moving
masses to point masses is simple. It is purely mathematical and conceptual. Newton correctly observed that the best we can do is to consider a
physical body as a point object with size of a geometric point. Without this
assumption, how can one define motion mathematically? With this assumption
Newton found a solid mathematical representation of motion in a space as a
function from time to space. With this he obtained the legitimate concept of
speed as the first order derivative of the motion function and the legitimate
concept of acceleration as the first order derivative of the speed function.
Without all of this basics, we have no theory of motion upon which to
build dynamics. Mathematics is an essential part of physics and not just
a language for physics. }

\textrm{Moreover, for dynamics, we have yet another important reason to
reduce a mass to a point mass. It is because force is a vector, a pointed
entity. So, the only entity to which we can exert a force is a point object
(mass). }

\subsection{Acceleration-induced gravitational field}

\textrm{There are some more issues to be discussed regarding the
``gravitational field" that Einstein introduced to a space under
acceleration. As discussed in the foregoing, the concept of any force field,
in general, violates the action-reaction law, which in turn violates the principle
of relativity. Moreover, the gravitational field that Einstein introduced to an
accelerating space is a force field which has no source for the
gravitational forces spreading all over the space. This is yet another
violation of the third law in a different sense. The ``uniform gravitational
field" near the surface of our planet is a gravitational
field in ``approximation". }

\textrm{More importantly, the ``force field", whatever it may be, that
Einstein introduced to frames under acceleration is not gravitational at
all. Gravitational fields are to be the modal representation of the effect
of gravitational force created by Newton's law of gravity. So, it is not a
uniform field. \textrm{A possible argument to counter that would be that near the surface of
the Earth the gravitational field is ``almost" uniform. This is however not acceptable in
precise science such as theoretical physics. Almost uniform is not uniform. The Universe is not approximate. }}

\textrm{It appears that the idea of associating ``acceleration" with
``gravitational force" comes from the old concept of ``aether" by Descartes.
Descartes wrongly considered a spaceship that contains an object which is under acceleration and identified the fictitious (inertial) force with gravitational force. As the name ``fictitious force" clearly indicates, such
a force is just an imaginary force that in fact does not exist. It is not
that the acceleration exerts such a force but as the object in the
accelerating frame is not a part of the frame (spaceship) it appears that
everything in the spaceship moves with acceleration relative to the object. Hence, it is not a real force. It is a fictitious force. }

\subsection{Red shift and energy issue}

\textrm{Einstein assumed a laboratory that is free falling under the
gravitational force. Assume we emit a light beam upward from the floor to
the ceiling. Due to the acceleration, by the time the light reaches the
ceiling, the ceiling is moving faster than the source on the floor was when
the light beam left it. In other words, the receiver at the ceiling is
approaching the source (where it was when the light left). 
\emph{Therefore the
observer in the lab will notice the blue shift due to the Doppler effect.} This will make this observer
notice the downward acceleration. This contradicts the equivalence principle
which states that a free falling body will not notice its free falling. So,
Einstein postulated that there must be a red shift due to the light moving
upwards against the gravitational force to offset this blue shift.
Unfortunately, it is not the observer in the lab who sees that the ceiling
is moving towards the floor. It is an observer outside of the falling lab
who will see that the ceiling is falling towards where the floor used to be.
So, the observer inside the lab will not observe the blue shift. This is how
Einstein obtained the red shift effect.}

\textrm{The situation is rather complex and we have to consider many
elements involved in this apparently simple thought experiment. This is an
instance of the confusion coming from the ambiguity that a rest point in a
frame F1 is also a moving point in a reference frame F2 that is
moving relative to F1. The problem which haunted the special theory of
relativity, which we presented as the ``power pole-power line paradox" in the context of time
dilation and length contraction, came back to haunt the general theory of
relativity, in the context of accelerating frames this time. As inertial
frames and accelerating frames are both moving frames, the same problem affects 
both of them. So the equivalence principle which is to reduce an
accelerating frame to an ``inertial frame" with induced gravitational field 
has no capability to avoid this fatal contradiction. }

\textrm{Moreover, the connection between relativistic dynamics and the general
theory of relativity is not clear at all. Are they equivalent? Both of them
are supposed to be the relativity theory of dynamics dealing with more than
relativistic kinematics, which is known as the special theory of relativity.
We know the former is inconsistent as it violates the limitation to non-accelerating frames, which is badly needed to avoid the contradiction against the third law of
dynamics, an absolutely essential assumption for any dynamics. }

\textrm{Putting aside the problem associated with the inconsistency of the
relativistic dynamics, there are some more lessons to be learned from using
inconsistent theories such as the special theory of relativity, relativistic
dynamics and the general theory of relativity. Neither the special theory of
relativity nor relativistic dynamics can handle energy. What about the
general theory of relativity, which is based upon the equivalence principle?
Unfortunately the answer is negative as well. As we have discussed at the beginning
of this section, the special theory of relativity claims that the addition
of speeds should not be }$v+v^{\prime }$\textrm{\ but }$v\oplus v^{\prime }.$ 
\textrm{This contradicts the second law of dynamics. }

\textrm{All of this clearly suggests that Newton's dynamics is a theory of
dynamics in the absolute frame only. Relativity is introduced not in the way
Galileo and Einstein did in their relativity theory. Relative motions are
defined as the difference of two absolute motions. All attempts to
relativize Newton's dynamics thus far have failed rather expectedly. }

\subsection{Centre of masses in general relativity theory}

\textrm{As discussed above, when we represent a constant speed motion in the 3D
Euclidean space as its graph, it becomes a straight line in the 4D
space-time. If the motion is an accelerating one, then the graph becomes a curved
line in the 4D space-time. Einstein represented gravitational force field,
which will cause accelerating motions, as the 4D space-time manifold so that
all motions caused by the gravitational field will appear as straight lines
(geodesics) in the manifold. Applying this idea to gravitational fields
created by a system of masses, Einstein obtained the so-called
gravitational field theory. In this theory the spatial distribution of
masses determines the 4D space-time manifold representing all possible
``motions".}

\textrm{According to the classical dynamics, when we have a system of masses
in a space, the mutual gravitational pull makes them converge to a single
point called the \textsl{centre of gravity}. \textsl{It is unclear how this process of convergence is dealt with in the gravitational field
theory of Einstein. }This\textsl{\ }question relates to the capacity of this
theory to express dynamic processes. }

\subsection{Light bend}

\textrm{Einstein studied the rest mass }$0$\textrm{\ photon under constant
acceleration. He considered a photon moving with speed }$c$\textrm{\ in the }%
$x$\textrm{-direction while it is in a frame that is under acceleration }$a$%
\textrm{\ in the }$y$\textrm{-direction. So, we have}

\[
x\prime =ct\qquad y\prime =-(at^{2})/2  \quad \quad(1) 
\]%
\textrm{If }$\theta $\textrm{\ is the angle made by a tangent of the light
ray to the }$x$\textrm{-axis, we have}

\[
tan(\theta )=-ax^{\prime }/c^{2}, 
\]%
\textrm{and we can assume that }$\theta $\textrm{\ is very small. So we have}

\[
\theta \doteqdot -ax^{\prime }/c^{2} \quad \quad(2) 
\]%
\textrm{But the GTR predicts otherwise, i.e.}

\[
\theta =-(3a/2c^{2})x^{\prime 2} \quad \quad(3) 
\]%
\textrm{\ Note that Eddington proved experimentally that the GTR's prediction }$(3)$%
\textrm{\ is correct.}

\textrm{It is clear that \textsl{all of this uses nothing but the kinematic
concept of acceleration and in kinematics there is no concept of mass}.
Putting aside the issue of inconsistency coming from the assumption of the
mass }$0$\textrm{\ point mass as discussed in section 3.1.3, the concept of
photon belongs to dynamics. The photon is a point mass of mass }$0$\textrm{.
Having no mass and having mass }$0$\textrm{\ are entirely different categories.%
\textsl{\ The problem with considering a point object whose mass is }}$0$%
\textrm{\textsl{\ is that the second law fails for the mass }}$0$\textrm{%
\textsl{\ point object. So,} \textsl{it makes no sense to say that the light
(trajectory of photon) bends due to the gravitational force of the Sun.} Thus,
Einstein's argument here diverted this difficulty by replacing the
``gravitational force upon mass }$0$\textrm{\ photon" with the ``reference
frame of photon accelerated by the Sun's gravitation". It is truly astounding
that this even worse confusion has never been detected till now. \textsl{The
reference frame of this photon cannot be accelerated unless there is some
mass stationary in it.} The second law was never meant to be applied to mass
zero objects.}

\textrm{\textsl{So,}} \textrm{\textsl{the only apparently appropriate
thing to say here is that the claim that light bends due to the acceleration of the
reference frame caused by the gravitation force is also false. In the end,
we do not know what is really happening here. If the curved 4D spacetime GTR
predicts the equation (3), then clearly something that went wrong in
the development of 4D spacetime GTR. }Unfortunately, Einstein's rest mass }$%
0 $\textrm{\ particle that  moves with speed }$c$\textrm{\ was brought into
physics with serious consequences. }

\subsection{``Induced gravitational field" revisited}

\textrm{As discussed in the foregoing, the concept of any force field in
general violates the action-reaction law, and in turn violates the
principle of relativity. Moreover, the gravitational field that Einstein
introduced to an accelerating space is a force field that has no ``external
source" creating the ``gravitational forces per unit mass" spreading all over
the space. This is yet another violation of the third law in a different
sense. The ``uniform gravitational field" }$g$\textrm{\ near the surface of
our planet is a gravitational field in ``approximation". It is
``not" a gravitational field. So, it is wrong to call Einstein's ``induced"
force field a gravitational force field.}

\textrm{To simplify this already confusing situation, let us put it this
way. In classical post-Newtonian dynamics, a field of
gravitational force applied to each location on a unit mass placed there was considered.
This concept itself fundamentally violates the action-reaction law and so  must
be abandoned. The situation here with Einstein's gravitational field created
by the empty reference frame under acceleration makes things even more
problematic. The former ignored the source mass that created the force field.
Here such a source mass to be ignored does not exist at all. }

We must stop identifying entirely different things in approximation as it was done in quantum field theory. 

Moreover, the force field that Einstein introduced to the frame of the spaceship is not
gravitational at all. Gravitational fields are to be the field
representation of the effect of gravitational force created by Newton's law
of gravity. So, it is not a uniform field.

\textrm{It was Minkowski's 4D spacetime theory of relativity which taught
Einstein that the motion line (geodesic) in the 4D spacetime bends in the
presence of acceleration. Combining this with the second law of dynamics,
which connects acceleration and force through mass, and the law of gravity,
Einstein concluded that through gravitational force masses distributed in
the 4D spacetime bend the 4D spacetime. Now the 4D space itself is
curved and the motion line is a straight line (geodesic) in this
curved 4D spacetime. So, there is no need for TD, LC and Lorentz transformation. But then what about Einstein's claim that Minkowski
spacetime is a local tangential space-time in the curved spacetime? }

\section{General theory of relativity (II)}

\subsection{General coordinate system}

\textrm{Judging from the argument to claim the equivalence principle, it
appears that Einstein was considering only local situations, such as a spaceship, as an accelerating frame. In this local setting it appears
possible to treat all accelerating frames as inertial frames. Without
knowing that the equivalence
principle is false, which we explain above, Einstein made a move to represent all accelerating
reference frames as ``local inertial frames" in which acceleration is
replaced by the induced gravitational field. \emph{This ill-fated idea lead
Einstein to consider the general coordinate system upon which all
accelerating frames are treated as local inertial frames with induced
gravitational field. }Hence, Einstein moved on to develop the concept
of general absolute reference frames to which we will turn in what follows. }

\textrm{Einstein assumed that the whole cosmos is occupied by a fluid whose
molecules are ``clocks" of any variety. This fluid can flow in any manner
except that it will be assumed that there is no turbulence, so that
neighbouring molecules always have almost equal ``speed" and the velocity of
the flow is a continuous function.}

\begin{remark}
\textrm{This means that Einstein assumes a universal time and a universal
space upon which clocks move.}
\end{remark}

\textrm{Each clock is allocated three coordinates }$(x_{1},x_{2},x_{3})$%
\textrm{\ in such a manner that:}

\begin{enumerate}
\item \textrm{No two clocks will have the same coordinates, and }

\item \textrm{Neighbouring clocks have neighbouring coordinates. Therefore,
coordinates are also continuous with respect to spatial displacement. }

\item \textrm{It is understood that the coordinate of each clock remains the
same through time. As time elapses at each clock, its readings are assumed to
increase but the rate of increase is not necessarily uniform as compared with
a local standard clock. No attempt is made to synchronize distant clocks.
 Neighbouring clocks are assumed to be ``sufficiently synchronized" so that the
clock readings are continuous with respect to spatial displacement. }

\item \textrm{The reading of a clock will be denoted by }$x_{0}.$\textrm{\ }
\end{enumerate}

\textrm{It is unfortunate that this paradigm is not possible for the following reasons: }

\begin{enumerate}
\item[a)] \textrm{No clock has any specific coordinate as it is not a point
object. }

\item[b)] \textrm{\ In the continuum, there is no such thing as a point next
to another point. So, the concept of a ``neighbouring point" to a point is
invalid. 
There is no such thing as a rational number next to a given
rational number. This is because in between any two rational numbers, we can
always find a rational number. This property is called the density of the
set of rational numbers. }

\item[c)] \textrm{As pointed out in \textbf{b}), there is no such thing as
the coordinate of a clock in this setting. Clocks are made by continuumly
many points. They are by themselves very complex physical
structures. In \textbf{a}) Einstein assumes the fluid of clocks and
yet in \textbf{c}) he says that the coordinate of each clock remains the same. }
\end{enumerate}

\textrm{More generally, the following further questions remain to be
answered. }

\begin{enumerate}
\item[(1)] \textrm{Upon what time and space the mechanics of such molecule
clocks is defined? Each clock is a physical system and so it is operating
in a spacetime that is not the same as the spacetime defined by the clock.
This is to say that the spacetime }$(x_{0},x_{1},x_{2},x_{3})$\textrm{\ does
not define the inside dynamics of the clock at }$(x_{0},x_{1},x_{2},x_{3})$%
\textrm{. Moreover, where is the clock which governs the spacetime in which
this clock operates? According to the general theory of relativity, the time
of this spacetime }$(x_{0},x_{1},x_{2},x_{3})$\textrm{\ and that of the
spacetime in which this clock operates are not the same and how much they
are synchronized depends upon the location of the clock that defines the
spacetime that defines the clock. This problem is related to a more
general problem associated with the instrumentalist view of time as clocks.
This view falls into the following vicious circle: The clock, which is
supposed to define time, must operate as a dynamical system upon some time
and space. Then how this time and space are supposed to be defined?}\emph{\ }

\item[(2)] It is a common understanding among researchers in ``dynamical
system theory" that time has a special status and is different from all other
coordinates of the system. This is in agreement with the idea of Newton in
his classical dynamics. Newton said that time, unlike other coordinates, has a 
natural flow that ``moves" forward only. This makes it impossible to
consider time as reading of clocks. Time is an entity that transcends
empiricism and operationalism. 

Once we
violate the most fundamental assumption on time, anything can happen and
relativity apparently made it happen.

\item[(3)] \textrm{Clocks are physical entities. There are at most countably
many clocks in this universe. No mater how closely we put clocks together,
we cannot form a continuum of clocks. No matter how one puts countably
infinite particles together, one will not create a mathematical continuum. This
is mathematically the same problem as the problem of photons, which are
supposed to exist for each frequency: as the frequency has a continuous
spectrum, there must exist uncountably infinite particles called photons.
Countably infinite points will never form real continuum. We need
continuumly many points to form mathematical continuum. }

\item[(4)] \textrm{What does it mean to be sufficiently synchronized? The
concept of synchronization presupposes external absolute time which
contradicts the concept of relativism. Here, we have to check time of each
clock at precisely the same moment in absolute time.}
\end{enumerate}

\subsection{Minkowskian local frame}

\textrm{Suppose that at a point }$\emph{P}$\textrm{\ in a ``gravitational field",
which is sea of infinitely many clocks, a freely falling non-rotational
(relative to distant stars) local inertial frame is ``constructed". \emph{We
further assume that the axioms of the special theory of relativity are valid
within this frame as it is supposed to be an inertial reference frame.} So,
we can set up a Cartesian coordinate system }$(\emph{Px,Py,Pz)}$\textrm{\ at
this point }$\emph{P}$\textrm{. Furthermore, we can distribute clocks over
the frame, all of them synchronized to the clock at }$P$\textrm{. }

\textrm{As we assumed that the universe is a sea of clocks that are not
overall synchronized, this implies that such a coordinate system }$(Px,Py,Pz)$%
\textrm{\ is not universal. It is a local coordinate system around }$P$%
\textrm{. }

\textrm{Using this frame and clocks, events that occur in the vicinity of }$%
\emph{P}$\textrm{\ over a suitably restricted time period, can be allocated
space-time coordinates }$(t,x,y,z)$\textrm{. }

\textrm{It is not quite clear why the time period must be restricted.}

\textrm{Now suppose that in this local inertial frame a pair of neighboring events have space-time coordinates }$(t,x,y,z)$\textrm{\ and }$%
(t+dt,x+dx,y+dy,z+dz).$\textrm{\ Then, if }$d(\tau )$\textrm{\ such that }

\[
(d(\tau ))%
{{}^2}%
=(dt)%
{{}^2}%
-(1/c)((dx)%
{{}^2}%
+(dy)%
{{}^2}%
+(dz)%
{{}^2}%
) 
\]%
\textrm{is Lorentz invariant, then it also is called the \emph{Minkowski
distance}. This serves as the correct metric on the 4D Minkowskian
spacetime. }

\begin{remark}
\textrm{As we will discuss later, Lorentz transformation
is irrelevant to theoretical physics as the claim by Lorentz that this
transformation maps wave equations to wave equations is false and so Einstein's
claim that all equational axioms of Maxwell are Lorentz invariant is false too. So,
Minkowski distance also is irrelevant to physics. Mathematical relevance of
such transformation is highly questionable either. }
\end{remark}

\textrm{There are some mathematical problems regarding this metric on the
``local" 4D spacetime. (1): It is not a topological metric that is used in topology. This is to say that Minkowski 4D space time is not a metric space. (2):}
\textrm{The Minkowski distance between two events that happen at the same
time is zero regardless of the 3D geometric distance between these two
events. (3)}: \textrm{Here, Einstein is assuming that in this ``freely
falling" local inertial frame, in which a gravitational field is induced
by the equivalence principle, all clocks are synchronized. The local inertial
frame must be accompanied by a gravitational field. So, all of these
clocks are under gravitational acceleration. How is it possible then that all of
these clocks are synchronized? Einstein also claims that all clocks
under acceleration slow down. Do they slow down uniformly? As the
acceleration is inertial ``locally", and time dilation is relative to the
inertial speed, this slowdown is not uniform at all.}\emph{\ }

\textrm{After all, as we have shown in the foregoing, the Minkowski distance
has no relevance to theoretical physics. It is mathematically irrelevant too
as it is not a topological metric. This clearly shows where relativity
theory should be placed in science. It is neither physics nor mathematics.
It appears that it belongs to its own category. }

\textrm{The local inertial frame is suitable only for the description of
very limited situations. For a larger scale (temporal, as well as spatial)
issues, it is necessary to use one of the general reference frames. If }$%
x_{i}$\textrm{\ are space-time coordinates relative to such a general frame,
transformations of the form }$x_{i}\prime =\pi (x_{0},x_{1},x_{2},x_{3})$%
\textrm{\ must exist relating }$(x_{0},x_{1},x_{2},x_{3})$ \textrm{to }$%
(t,x,y,z)$\textrm{\ [the local inertial frame] \ such that }

\[
t=\theta (x_{0},x_{1},x_{2},x_{3}),\;x=\pi
(x_{0},x_{1},x_{2},x_{3}),\;y=\psi (x_{0},x_{1},x_{2},x_{3}),\;z=\gamma
(x_{0},x_{1},x_{2},x_{3}). 
\]

\textrm{Then, if }$x_{i}$\textrm{\ are subjected to increments }$dx_{i}$%
\textrm{, the corresponding increments in }$t,x,y,z$ \textrm{will be given by%
}

\[
dx=(\partial (\theta )/\partial (x_{0}))dx_{0}+(\partial (\pi
)/\partial (x_{1}))dx_1 +(\partial (\psi )/\partial (x_{2}))dx_2 +(\partial (\gamma )/\partial (x_{3}))dx_3 
\]%
\textrm{etc. and substitution in equation of proper time interval }%
\[
(d(\tau ))%
{{}^2}%
=(dt)%
{{}^2}%
-(1/c)((dx)%
{{}^2}%
+(dy)%
{{}^2}%
+(dz)%
{{}^2}%
) 
\]%
\textrm{will result} \textrm{in an expression }$d\tau 
{{}^2}%
$\textrm{\ that is quadratic in the increments }$dx$\textrm{, i.e. whose
terms will either involve squares of the }$dx_{i}$\textrm{\ or the product of
two different }$dx_{i}$\textrm{. Thus}

\[
d\tau 
{{}^2}%
=\sum_{i=0}^{3}\sum_{j=0}^{3}(g_{ij})dx_{i}dx_{j}\qquad \qquad (R) 
\]%
\textrm{where the coefficients }$(g_{ij})$\textrm{\ will be the functions of 
}$x_{i}$\textrm{.}

\textrm{Now, a continuum in which the interval between neighbouring points is
given by a quadratic form like }$(R)$\textrm{\ is called a ``\emph{Riemannian
space}" and the quadratic form like }$(R)$\textrm{\ is called its \emph{%
metric}. Thus, the space-time continuum is a four-dimensional Riemannian
space whose interval is everywhere identified with the proper time interval
between neighbouring events in a local inertial frame.}

Here
are some issues to be discussed:

\begin{enumerate}
\item \textrm{\ What is important for physics is not that we use 4D
spacetime manifold of Riemann. What has been questioned here is the
relevance of such a mathematical structure to physics. This structure }$(R)$%
\textrm{\ is obtained by substitution }%
\[
dx=(\partial (\theta )/\partial (x_{0}))dx_{0}+(\partial (\pi
)/\partial (x_{1}))dx_1 +(\partial (\psi )/\partial (x_{2}))dx_2 +(\partial (\gamma )/\partial (x_{3}))dx_3
\]%
\textrm{etc.} \textrm{in the equation of proper time interval }%
\[
(d(\tau ))%
{{}^2}%
=(dt)%
{{}^2}%
-(1/c)((dx)%
{{}^2}%
+(dy)%
{{}^2}%
+(dz)%
{{}^2}%
). 
\]%
\textrm{As we have shown, this whole mathematical argument makes little
physical sense. A most serious flaw in all of this is that the theory of general relativity was obtained from the special
theory of relativity which is both mathematically and ontologically inconsistent. So, the general theory is false too. As discussed above, the special theory of relativity yields the relativistic
addition of speed which contradicts the addition of acceleration of a mass
which is governed by the second law of Newton. The equivalence principle,
which is the most fundamental assumption of the general theory of relativity,
is based upon the second law of dynamics. 
Einstein explains
the curved space in the general theory of relativity using a rotating disc;
the radius of the disk does not contract as the motion of the disk is not
along this direction. It comes under the influence of the length contraction
around the perimeter of the disk as the motion is in the direction of the
tangential speed of the point on the perimeter of the disk. There are two
errors in this argument. Firstly, the tangential speed of a circular motion is
not inertial, it is under acceleration and so, the length contraction
should not apply. Secondly, length contraction is false. }

\item More fundamentally, 
Einstein was clearly not aware of the difference between
a countably infinite and a continuum.
Cantor's diagonal argument clearly shows that there are more points in the
geometric continuum than the discrete collection of points. The Lebesgue
integral of the Weierstrass function over $[0,1]$ shows that the
geometric continuum has unimaginably more points than the ``space" of countably
many points has. For example, on the real number line almost all points are
irrational numbers. So, one cannot cover the entire global space with clocks
as there are only finitely many clocks. This makes the most fundamental
assumptions of Einstein's general theory of relativity untenable. There is
no such thing as the ``global spacetime" prescribed by Einstein. define 
neither differentiation nor integration.

\item \textrm{Also, Einstein's description of the clocks used to define the
global spacetime is off.} \textrm{To begin with, it must be required that the
neighbouring clocks are of infinitesimal distance and the time difference
between each pair of neighbouring clocks must be infinitesimally small. Otherwise we
cannot use calculus to calculate on such a structure. Physically, it is
impossible to make enough clocks to do this and place them in the way
expected, as we stated above. 
As an infinitesimal means a number that comes in between }$0$\textrm{\ and any positive number, this method is ontologically
untenable. Clearly Einstein was unaware of what infinitesimals were as this
concept was articulated only later in 1960's by Abraham Robinson. Even in pure
mathematics, the work of Robinson is understood only by a small number of
mathematical logicians. The general theory of relativity was developed before
the development of Robinson's infinitesimal calculus and so it is
understandable that the Hilbert school of mathematics and physics did not
have the concept of infinitesimals. By the time Robinson's work came out, the
separation between pure mathematics and theoretical physics became material
and communication between these two communities became almost non-existent. }

\item In short, contrary to what Einstein proposed, the universe cannot be a sea of clocks. 
 
In addition to this topological
problem the general theory of relativity suffers, there is an even more
fundamental issue of logical deficiency in the idea of the general
reference frame which is the sea of clocks. Clocks are physical entities and
it requires physics to make them. One cannot use clocks to define clocks at
the pain of vicious circle. So, there is no such thing as metaphysical
clocks though time is certainly a metaphysical entity, as Newton thought. It
was relativity theory, the special and the general, which tried to use empirical
clocks that lead the world of physics to the current confusion about time.

Logically speaking, modern physics started with the
wrong idea of what is time. Contrary to the special theory of relativity,
time cannot be defined in terms of speed as speed is defined in terms of
time. And as we have discussed here, the universe is not a sea of clocks
contrary to the general theory of relativity. From the combination of these wrong assumptions, it is expected that we ended up with 
questioning what time is. 

Our understanding of
time as in relativity theory is completely wrong. 

\end{enumerate}

We have 
shown that the general theory of relativity came from the special theory of relativity, which is false. Therefore, the general theory of relativity is also false. Any theory which contains an inconsistent
theory is inconsistent. 

\subsection{Geodesics}

\textrm{When we express a linear function of one variable on a 2D space, then
the function's graph becomes a straight line. The coefficient of the first
order variable is the slope of the line. This idea was extensively exploited
by train companies to visualize train operation on a 2D space where
one coordinate is the time coordinate and the other coordinate is the location
coordinate expressed at the distance from the origin station. It is called
``operational diagram", which pure mathematicians came to treat as a 4D
spacetime. In the 4D space time all
constant speed 3D motions should be just straight lines and the slope of the
line is the constant speed 3D motion. }

\textrm{So, in 4D spacetime geometry of 3D motions, the ``Euclidean geometric
distance" between two points} $P_{1}(t_{1},x_{1},y_{1},z_{1})$ \textrm{and} $%
P_{2}(t_{2},x_{2},y_{2},z_{2})$ \textrm{in the 4D spacetime is:} 
\[
\overline{P_{1}P_{2}}=\sqrt{%
(t_{1}-t_{2})^{2}+(x_{1}-x_{2})^{2}+(y_{1}-y_{2})^{2}+(z_{1}-z_{2})^{2}}. 
\]%
\textrm{The slope of the line segment }$P_{1}P_{2}$ \textrm{is given as} 
\[
\left( (x_{1}-x_{2})/(t_{1}-t_{2}),\quad (y_{1}-y_{2})/(t_{1}-t_{2}),\quad (z_{1}-z_{2})/(t_{1}-t_{2})\right) . 
\]

\textrm{In general theory of relativity, this changes: the distance, as par
Minkowski, between }$P_{1}$\textrm{\ and }$P_{2}$\textrm{\ is} 
\[
\widetilde{P_{1}P_{2}}=\sqrt{(t_{1}-t_{2})^{2}-(1/c^{2})\left\{
(x_{1}-x_{2})^{2}+(y_{1}-y_{2})^{2}+(z_{1}-z_{2})^{2}\right\} }. 
\]%
\textrm{It can be shown that this distance is smaller than any other
``Euclidean distance" along any curve connecting }$P_{1}$\textrm{\
and }$P_{2}$\textrm{\ in the 4D space. It goes as follows. Clearly, }%
\[
\widetilde{P_{1}P_{2}}<\overline{P_{1}P_{2}}. 
\]%
\textrm{But for any path }$l_{P_{1},P_{2}}$\textrm{\ between }$P_{1}$\textrm{%
\ and }$P_{2}$\textrm{, }%
\[
\overline{P_{1}P_{2}}\leq \overline{l_{P_{1},P_{2}}} 
\]%
\textrm{where }$\overline{l_{P_{1},P_{2}}}$\textrm{\ is the Euclidean length
of }$l_{P_{1},P_{2}}.$

\textrm{From this, relativity theory concludes that the 4D spacetime with Minkowski
metric is not a Euclidean space but a Riemannian space that is curved.} 
\textrm{There are some unclear issues to be addressed here.}

\begin{enumerate}
\item \textrm{In Euclidean geometry, we define the distance between two
points P and Q using the Pythagorean formula and we show that the length of any
path connecting the two points P and Q is longer or equal to the distance
between them. We have not seen how to define the length of a curved path
between P and Q in Minkowski 4D spacetime. }

\item \textrm{Putting this purely mathematical issue aside, the only
reasoning for admitting }$\widetilde{P_{1}P_{2}}$\textrm{\ , the Minkowski
distance between }$P_{1}$\textrm{\ and }$P_{2}$\textrm{, is because this
distance is invariant under Lorentz transformation. There are two questions
to be answered here however. (1) Why Lorentz transformation is the issue in the
general theory of relativity? This transformation belongs to the special
theory of relativity. (2) This transformation is said to be essential for
the special theory of relativity as it is said to preserve all wave equations
and axioms of Maxwell's electromagnetic field theory. (3) We have shown that
this claim is false in our paper ``\emph{Reference Frame Transformations
and Quantization}". }
\end{enumerate}

\textrm{Nonetheless ,Einstein continued to develop his theory as follows:
Under the equivalence principle,} 
\[
d\tau 
{{}^2}%
=\sum_{i=0}^{3}\sum_{j=0}^{3}(g_{ij})dx_{i}dx_{j}=(dt)%
{{}^2}%
-(1/c)((dx)%
{{}^2}%
+(dy)%
{{}^2}%
+(dz)%
{{}^2}%
).\qquad 
\]%
\textrm{From this, Einstein concludes that the free fall in the gravitational
field is a geodesics [in the 4D spacetime]. This is called the ``\emph{geodesic
principle}". Under this principle we obtain equations of motion for bodies
falling freely in a gravitational field. }

\textrm{Using Riemannian geometry for the ``general metric"} 
\[
d\tau 
{{}^2}%
=\sum_{i=0}^{3}\sum_{j=0}^{3}(g_{ij})dx_{i}dx_{j} 
\]%
\textrm{we can show the following ``general equation of a geodesic"}%
\[
\frac{d}{d\tau }\left( \sum_{j=0}^{3}g_{ij}\frac{dx_{j}}{d\tau }\right) =%
\frac{1}{2}\sum_{j=0}^{3}\sum_{k=0}^{3}\frac{\partial g_{ik}}{\partial
x_{i}}\frac{dx_{j}}{d\tau }\frac{dx_{k}}{d\tau }\qquad (i=0,1,2,3). 
\]%
\textrm{For the metric} 
\[
(d(\tau ))%
{{}^2}%
=(dt)%
{{}^2}%
-(1/c)((dx)%
{{}^2}%
+(dy)%
{{}^2}%
+(dz)%
{{}^2}%
), 
\]%
\textrm{these equations reduce to}%
\[
\frac{d^{2}x}{d\tau ^{2}}=\frac{d^{2}y}{d\tau ^{2}}=\frac{d^{2}z}{d\tau ^{2}}%
=\frac{d^{2}t}{d\tau ^{2}}=0. 
\]%
\textrm{This is equivalent to} 
\[
x=\frac{dx}{d\tau }t+a,\quad y=\frac{dy}{d\tau }t+b,\quad z=\frac{dz}{d\tau }%
t+c\quad 
\]%
\textrm{where} $a,b,c$ \textrm{are constants}.

\begin{remark}
\textrm{According to the general theory of geodesics, light coming from a
distant star passing near our sun has a geodesic (4D) which bends near the
Sun due to the gravity of the Sun. As it is a 4D bending, we cannot graphically
express this bending. But when we omit the time bending, the light path
bends in our 3D space. This is what we see in science museum exhibitions
everywhere. Then we have a problem to think about. This 3D bending is a
phenomenon that takes place in our 3D Euclidean space. This can be observed
only from the outside of our 3D space. As we are inside the 3D space, in theory, we
will be unable to observe this bending of light path. This was precisely
what George Gamow warned us about. Gamow basically said that unless we are in the position
of Newton out of the universe observing, we will not observe this bending.
This valid question was never answered scientifically. The late Prof. Marmet
presented a classical explanation of this observation using no bending space
but bending light in unbent space interpretation. It was ignored. This means
that we still do not know if space really bends as Einstein predicted due to
the gravitation of Earth. }
\end{remark}

\subsection{Einstein's equation of gravitation}

\textrm{The discussion above shows that }$g_{ij}$\textrm{\ determines the
motion (geodesic) in general relativity theory. Given the energy-momentum
tensor }$T_{ij}$\textrm{\ which describes the distribution of mass, energy
and momentum of the system, Einstein's equations of gravitation yield
corresponding }$g_{ij}$\textrm{\ enabling one to calculate the geodesics of the
system. It was Schwarzschild who first obtained an exact solution of
Einstein's equations for a spherically symmetric field. He used the
solutions to calculate the motion of a planet in the field of its sun. }

\textrm{Putting aside the problems of the general theory of relativity as
discussed above, this result of Einstein and Schwarzschild poses 
some issues to consider. The question is what this momentum-energy
tensor is about. There are several issues to be cleared.}

\begin{enumerate}
\item \textrm{The momentum-energy relation is a problem. The former is a
predicative concept but the latter, as the potential to do work, is a modal
concept, and so, connecting them in the same category is not the
right thing to do. }

\item \textrm{Logically speaking waves in wave mechanics have no momentum.
This is because momentum is the product of a mass and its speed. In wave
mechanics, no mass moves in the direction of the wave. What is moving
towards the direction of the wave is the local vibration of the medium. So,
waves have no momentum. When it comes to energy, as the work needed to
accelerate from }$m0$\textrm{\ \ to }$mv$\textrm{\ is not necessarily }$%
(1/2)mv^{2},$ \textrm{the concept of kinetic energy is false. The work
needed for this acceleration depends upon how we accelerate from }$m0$ 
\textrm{to} $mv.$

\item \textrm{So, the energy-momentum relation does not represent the state of a
physical system properly. Then how can the energy-momentum
tensor describe the physical system properly? }

\item \textrm{Is the energy-momentum relation as energy-momentum tensor used
here classical or relativistic?}\ \textrm{If it is relativistic, the entire
argument by Einstein becomes viciously circular. If not, then how does it relate to the claim that the classical physics is invalid. This is a 
contradiction, is it not? 
Classically we
have some tension between momentum and energy. Moreover, the relativistic
energy-momentum relation is also false. It is because }$e=mc^{2}$ \textrm{is
false as we discussed earlier in this paper. This equation came from the false
assumption that the }$v$\textrm{\ in the gamma factor in }$m$\textrm{\ is
time dependent, which is not allowed in relativity theory. Considering that
the concept of energy is not a physical reality but a modality and understanding that, as philosophy asserts, the modality and reality are of different
category, it is astounding that the momentum-energy tensor plays the most
fundamental role in general relativity theory. 
Mathematics is not just a language for physics. It
is the only way to understand physical nature around us. It is the most
articulate way of thinking correctly. }

\item \textrm{What is the most fundamental issue here is that the practice
of \emph{bootstrapping} classical mechanics to relativistic mechanics is not
a legitimate thing to do. }
\end{enumerate}

\section*{References}
\vspace*{15pt}
\nocite{*}
\bibliography{references-LAoRT}

\begin{thebibliography}{10}

\bibitem{bacon:1620}
F.~Bacon.
\newblock {\em Novum Organum}.
\newblock 1620.

\bibitem{bennett:1971}
J.~Bennett.
\newblock {\em Locke, Berkeley, Hume: Central Themes}.
\newblock Oxford University Press, 1971.

\bibitem{bergmann:1942}
P.~G. Bergmann.
\newblock {\em Introduction to the Theory of Relativity}.
\newblock Prentice Hall, 1942.

\bibitem{berkeley:1710}
G.~Berkeley.
\newblock {\em A Treatise Concerning the Principles of Human Knowledge (Part
  I)}.
\newblock 1710.

\bibitem{debroglie:1925}
L.~De~Broglie.
\newblock Recherches sur la th\'{e}orie des quanta (on the theory of quanta).
\newblock {\em Ann. de Physique}, 10, 1925.

\bibitem{dingle:1972}
H.~Dingle.
\newblock {\em Science at the Crossroads}.
\newblock Martin Brian and O'Keeffee, 1972.

\bibitem{einstein:1905}
A.~Einstein.
\newblock Zur elektrodynamik bewegter k\"{o}rper.
\newblock {\em Annalen der Physik}, 332:10, 1905.

\bibitem{engelhardt:2014}
W.~Engelhardt.
\newblock Phase and frequency shift in a michelson interferometer.
\newblock {\em Physics Essays}, 27, 2014.

\bibitem{heisenberg:1930}
W.~Heisenberg.
\newblock {\em The Physical Principles of the Quantum Theory (Translators
  Eckart C, Hoyt F C)}.
\newblock Dover, 1930.

\bibitem{kanda:2007}
A.~Kanda and P.~Apostoli.
\newblock Inconsistency of some formal theories.
\newblock In {\em presented at the Meeting of Exact Philosophers at University
  of British Columbia}, 2007.

\bibitem{kanda:2010}
A.~Kanda and P.~Apostoli.
\newblock Logicism v.s empiricism.
\newblock In {\em International Seminar on the structure in Cosmology and
  Theoretical physics, Finish Physics Foundation Society and Finish Society for
  Natural Philosophy, Helsinki}, 2010.

\bibitem{kanda:2019}
A.~Kanda, R.~Wong, and M.~Prunescu.
\newblock Reference frame transformations and quantization.
\newblock In {\em invited talk at Physics Beyond Relativity Conference, Praha,
  Czech Republic}, 2019.

\bibitem{kuhn:1962}
T.~S. Kuhn.
\newblock {\em The Structure of Scientific Revolutions}.
\newblock The University of Chicago Press, 1962.

\bibitem{lawden:1985}
D.~F. Lawden.
\newblock {\em Elements of Relativity Theory}.
\newblock Wiley, 1985.

\bibitem{lorentz:1904}
H.~A. Lorentz.
\newblock Electromagnetic phenomena in a system moving with any velocity
  smaller than that of light.
\newblock In {\em Proceedings of the Royal Netherlands Academy of Arts and
  Sciences}, 1904.

\bibitem{maxwell:1865}
James~C. Maxwell.
\newblock A dynamical theory of the electromagnetic field.
\newblock {\em Philosophical Transactions of the Royal Society of London},
  155:459--512, 1865.

\bibitem{miller:1970}
N.~Miller and S.~C. Ashby.
\newblock {\em Principles of Modern Physics}.
\newblock Holden Day, 1970.

\bibitem{minkowski:1907}
H.~Minkowski.
\newblock Das relativit\"{a}tsprinzip.
\newblock {\em Annalen der Physik}, 352:927--938, 1907.

\bibitem{robinson:1966}
A.~Robinson.
\newblock {\em Non-standard analysis}.
\newblock Princeton Landmarks in Mathematics. Princeton University Press, 1966.

\bibitem{schroedinger:1926}
E.~Schr\"{o}dinger.
\newblock An undulatory theory of the mechanics of atoms and molecules.
\newblock {\em Physical Review}, 28:1049--1070, 1926.

\bibitem{sells:1973}
R.~T. Sells and R.~L. Weidner.
\newblock {\em Elementary Modern Physics}.
\newblock Allyn and Bacon, 1973.

\bibitem{suntola:2011}
T.~Suntola.
\newblock {\em Dynamic Universe}.
\newblock Physics Foundations Society, 2011.

\end{thebibliography}


\end{document}